\documentclass[onecolumn,12pt]{IEEEtran}

\usepackage{algorithm,algorithmic,amsbsy,amsmath,amssymb,epsfig}

\DeclareMathAlphabet{\mathpzc}{OT1}{pzc}{m}{it}
\usepackage{times,amsmath,epsfig}
\usepackage{cite}
\usepackage{graphicx}
\usepackage{psfrag}
\usepackage{subfigure}
\usepackage{url}
\usepackage{stfloats}
\usepackage{amsmath}
\usepackage{amssymb}
\usepackage{array}
\usepackage{array}

\newcommand{\beq}{\begin{equation}}
\newcommand{\eeq}{\end{equation}}
\newcommand{\bitm}{\begin{itemize}}
\newcommand{\ba}{\begin{array}}
\newcommand{\ea}{\end{array}}
\newcommand{\eitm}{\end{itemize}}
\newcommand{\beqn}{\begin{eqnarray}}
\newcommand{\eeqn}{\end{eqnarray}}
\newcommand{\beqno}{\begin{eqnarray*}}
\newcommand{\eeqno}{\end{eqnarray*}}
\newcommand{\bma}{\begin{displaymath}}
\newcommand{\ema}{\end{displaymath}}
\newcommand{\bnu}{\begin{enumerate}}
\newcommand{\enu}{\end{enumerate}}
\newcommand{\bce}{\begin{center}}
\newcommand{\ece}{\end{center}}
\newcommand{\btb}{\begin{tabular}}
\newcommand{\etb}{\end{tabular}}

\begin{document}
%
%
%
%

\title{Game-theoretic Resource Allocation Methods for Device-to-Device (D2D) Communication}

\author{Lingyang Song, Dusit Niyato, Zhu Han, and Ekram Hossain
\thanks{L. Song is with the School of Electronics Engineering and Computer Science, Peking University, China. D. Niyato is with School of Computer Engineering, Nanyang Technological University, Singapore. Z. Han is with the Electrical and Computer Engineering Department, University of Houston, USA. E. Hossain is with the Department of Electrical and Computer Engineering, University of Manitoba, Canada.}
}

\maketitle

\thispagestyle{empty}
\begin{abstract}
Device-to-device~(D2D) communication underlaying cellular networks allows mobile devices such as smartphones and tablets to use the licensed spectrum allocated to cellular services for direct peer-to-peer transmission. D2D communication can use either one-hop transmission (i.e., \emph{D2D direct} communication) or multi-hop cluster-based transmission (i.e., in \emph{D2D local area networks}). The D2D devices can compete or cooperate with each other to reuse the radio resources in D2D networks. Therefore, resource allocation and access for D2D communication can be treated as games. The theories behind these games provide a variety of mathematical tools to effectively model and analyze the individual or group behaviors of D2D users. In addition, game models can provide distributed solutions to the resource allocation problems for D2D communication. The aim of this article is to demonstrate the applications of game-theoretic models to study the radio resource allocation issues in D2D communication. The article also outlines several key open research directions.
\end{abstract}


\section{Introduction}
As more and more new multimedia rich services are becoming available to mobile users, there is an ever-increasing demand for higher data rate wireless access. As a consequence, new wireless technologies such as LTE~(Long Term Evolution)/LTE-Advanced and WiMAX have been introduced. These technologies are capable of providing high speed, large capacity, and guaranteed quality-of-service (QoS) mobile services~\cite{Song2010}. With technology evolution of cellular networks, new techniques, such as small cells, have been also developed, which are able to improve network capacity by reducing cell size and effectively controlling interference. However, most attempts still rely on the centralized network topology, which requires mobile devices to communicate with an evolved Node B~(eNB) or access point~(AP). Such a centralized network topology can easily suffer from congestion by a large number of communicating devices. Also, the eNB and AP may not have complete information about transmission parameters among devices, which is required to optimize the network performance.

As an alternative, the concept of device-to-device (D2D) communication has been recently introduced to allow local peer-to-peer transmission among mobile devices bypassing the eNB and AP~\cite{Doppler2009}. Specifically, besides cellular operation, where the user equipments~(UEs) can be served by the network via the eNB in the LTE system, some UEs may communicate with each other directly over direct links for proximity-based services~\cite{3GPP}. The UE with D2D connections (i.e., a D2D user) is loosely controlled by the eNB. In particular, the eNBs can control the radio resource allocation for the cellular and the D2D links (i.e., the link between UE and eNB and the link between UEs, which we will refer to as cellular and D2D users, respectively). Also, the eNBs can set constraints on the transmission parameters (e.g., transmit power) of D2D users. The purpose of the constraints is to limit the interference experienced by the cellular users~\cite{WangF2013} and satisfy their quality-of-service (QoS) requirements.

\begin{figure}[h]
\begin{center}
$\begin{array}{c} \epsfxsize=4.6 in \epsffile{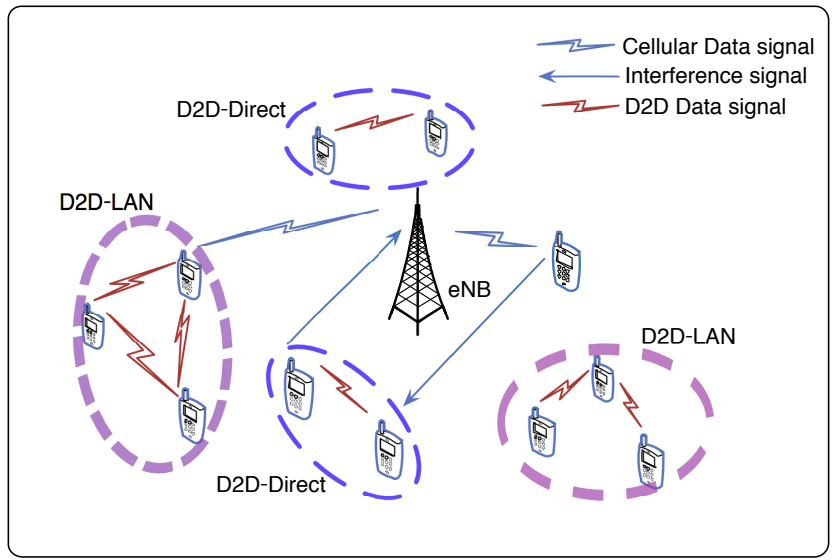} \\ [-0.2cm]
\end{array}$
\caption{D2D communication underlaying cellular networks including cellular communication, D2D direct, and D2D LAN.}
\label{f:sysmodel}
\end{center}
\end{figure}

In this article, we classify D2D communication into two categories: \emph{D2D direct} and \emph{D2D local-area network}~(D2D LAN). D2D direct simply refers to the conventional one-hop (one D2D pair) communication~\cite{Doppler2009}. In multi-hop D2D LAN, network-controlled smart devices can realize cluster-based communication in an ad hoc manner, and meanwhile work over the licensed band to achieve maximal flexibility and performance in a multi-cell scenario. Fig.~\ref{f:sysmodel} shows a typical single cell scenario with multiple users consisting of conventional cellular communication, one-hop D2D direct transmission, and D2D LAN for group communication.


Researchers and wireless engineers postulate that D2D communication will become a key feature supported by the next generation cellular networks. Based on peer-to-peer data transmission with loosely controlled radio resource allocation, the D2D communication provides the following advantages:
(1)	extended coverage~\cite{Pei2013},
(2)	offloading from cellular networks~\cite{Yang2013},
(3)	improved energy efficiency,
(4)	increased throughput and spectrum efficiency~\cite{Phunchongharn2013}, and
(5)	create opportunities for new services through mobile social networks and vehicular adhoc networks~\cite{Wang2013}.
D2D communication is allowed to reuse the same spectrum with cellular services as an underlay, thus improving the system throughput~\cite{Yu2011,Xu2012}. With spectrum sharing, although D2D communication can improve spectral efficiency and gain in system capacity, it also causes interference to the cellular network users. Therefore, methods for efficient interference management and coordination must be developed to achieve the target performance levels for both the cellular and D2D users which maximizes the spectrum utilization.


To restrict co-channel interference, the problem of power control for UEs (i.e., D2D UEs or D2D users) has been studied in the literature~\cite{Doppler2009}. To further improve the gain from intra-cell spectrum reuse, the problem of pairing the cellular and D2D users for sharing the same radio resources was studied in~\cite{Janis_VTC}. \cite{Yu_TWC} provided an analysis on optimum resource allocation and power control between the cellular and D2D connections. Also, \cite{Yu_TWC} evaluated the performance of the D2D underlaying system in both a single cell scenario and the Manhattan grid environment. All these work along with many other related work considered the spectrum reuse of cellular and D2D users from an optimization perspective. That is, all the users must agree on the same single objective. Also, all the users must be fully controlled through extensive signaling to achieve such an objective.


Generally, for both D2D direct and D2D LAN, cellular and D2D users may be of self-interest to maximize their own benefits through cooperation and/or competition in the underlaying D2D network. This will require the cellular users, D2D users, and the base station (i.e., an eNB in an LTE-A network) to solve distributed decision problems. Consequently, there is a need to develop suitable solutions to these problems for efficient D2D communication. In particular, solution approaches are required which will provide the following:
\begin{itemize}
	\item {\em Theoretical foundation:} a theoretical basis to analyze interactions in multiple player systems,
	\item {\em Distributed operation:} wireless nodes should be able to make their decisions independently using the local information with small amount of signaling overhead, and
	\item {\em Mechanism design:} the parameters of the models can be designed (or varied) such that they lead the independent and self-interested wireless nodes toward a system-wide desirable outcome.
\end{itemize}

Game theory offers a set of mathematical tools to study complex interactions among interdependent rational players and predict their choices of strategies. Therefore, game theory becomes a suitable tool to model and investigate the resource allocation problem for D2D communication. Also, game theoretic models can be developed to obtain solutions for channel assignment, power control, and cooperation enforcement among D2D users. In contrast with the existing game-theoretical works on wireless resource allocation, network controlled D2D communication allows more flexible frequency reuse, and enables more types of applications, specially for D2D LAN, such that requiring systematic studies of game models. In this article, we demonstrate how game theory can be applied to study the resource allocation problem for D2D communication. In addition to reviewing the existing approaches, we also highlight some potential research directions toward using game theory to address other open problems in D2D communication.


\section{Basics of D2D Communication}\label{sec:D2D basics}

In this section, we give an overview of the D2D direct and D2D LAN communication scenarios. For both types of these communications, the D2D users can access the spectrum in two different ways:

\begin{itemize}
 \item {\em Overlay spectrum sharing:} This approach can completely eliminate intra-cell interference between cellular services and D2D communication by dividing the licensed spectrum into two parts through orthogonal channel assignment. One part is a set of channels to be used by the cellular users while the other part will be used by the D2D users.
 \item {\em Underlay spectrum sharing:} This approach allows D2D users to use the same spectrum simultaneously with cellular users, increasing spectrum utilization and efficiency. However, the interference among users must be carefully controlled and avoided.
\end{itemize}
The overlay approach is easier to realize. However, it results in low spectrum utilization and efficiency as the different parts of spectrum cannot be flexibly and opportunistically accessed by cellular and D2D users. On the other hand, the underlay approach requires relatively larger signaling overhead and controlling information. Nevertheless, it can achieve better system performances due to the frequency reuse with cellular users.

In this article, we mainly focus on the underlay approach. Fig.~\ref{f:sysmodel} shows the network with two tiers, i.e., D2D and cellular tiers, and interference in such a network can be classified as follows:
\begin{itemize}
 \item {\em Cross-tier interference:} This is the interference between users in different tiers. The transmission of an aggressor (e.g., a D2D user) interferes that of a victim (e.g., a cellular user).
 \item {\em Co-tier interference:} This is the interference between users in the same tier. The transmission of the aggressor (e.g., a D2D user) interferes that of the victim (e.g., another co-channel D2D user).
\end{itemize}
Therefore, to reduce the appearance of dead zones (i.e., the zones that users cannot transmit data due to strong interference) within the cellular networks and successfully enable D2D communication, interference avoidance, randomization, or cancelation techniques must be effectively applied. For this, the spectrum allocation needs to consider the interests of  the cellular network (operator) and mobile users (cellular users and D2D users).

Unlike traditional resource allocation in orthogonal frequency-division multiple access (OFDMA) networks, the resource allocation in D2D involves the assignment of a D2D user to proper resource blocks~(RBs), which affects the co-channel interference~(CCI) from the co-channel cellular user. Specifically, the network radio resource allocation can be performed in the following ways:
\begin{itemize}
 \item {\em Local radio resource allocation:} The radio resource allocated to cellular users is considered to be fixed, while that of D2D users can be adjusted.
 \item {\em Global radio resource allocation:} The radio resource of both cellular and D2D users can be controlled jointly.
\end{itemize}

Distributed radio resource allocation algorithm can be designed based on game theory~\cite{Han2011}. In particular, a game theory model can be developed to study the interactions between a cellular network, which has some commodity or service to sell (in this example, the commodity is radio spectrum), and a number of cellular and D2D users interested in obtaining the service so as to optimize their objective functions, respectively. To this end, auction theory offers such a set of mathematical tools to design the parameters of the whole network, so as to efficiently allocate the radio resources for D2D direct communication. One such model will be elaborated later in this article.


With regards to D2D LAN, the network-controlled mobiles can perform group communication, and thus, realize various functionalities based on specific application scenario. Similar to D2D direct, these mobiles in the D2D LAN can also work as an underlay to cellular networks for spectrum reuse. The scenarios are as follows:
\begin{itemize}

 \item \emph{Group communication:} When a large number of similar requests are received by the eNB, the D2D LAN can be used to efficiently offload data. For example, in stadium or concert networks, when many mobile users request for the same content, some UEs as `{\em seeds}' can be first selected to obtain the complete information from the eNB, and then these seeds can share the data with the rest of the mobiles (Fig.~\ref{seed}).

 \begin{figure}[h]
\begin{center}
\includegraphics[width=4.0in]{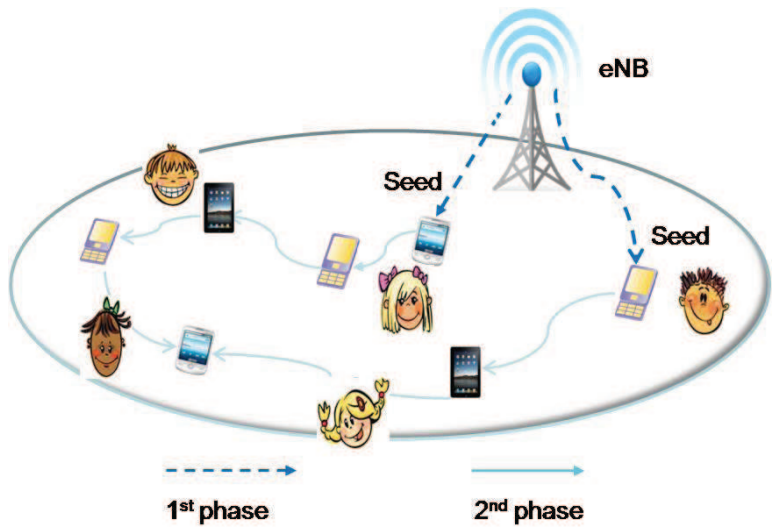}
\caption{Popular content downloading in hotspot areas, such as concert and stadium networks.}
\label{seed}
\end{center}
\end{figure}

 \item \emph{Multi-hop relay communication:} When some users are out of the coverage of the eNB, the mobiles in the D2D LAN can serve as relays for completing the file delivery among mobiles.

  \item \emph{Collaborative smartphone sensing:} Smartphones have the capability of environment sensing and the sensed data can be collaboratively aggregated to some `sink' UEs and then transmitted to the eNB.
\end{itemize}

Apparently, for D2D LAN, it is essential to form clusters among D2D users, and this problem can be formulated as a \emph{cooperative game}. Specifically, a \emph{coalition formation} algorithm which converges to a stable solution can be used with a distributed solution. The details will be elaborated later in this article.

\section{Game Theory-Based Resource Allocation for D2D Direct Communication}

In this section, we discuss different game theoretic models to solve the resource allocation problems for D2D direct communication. Also, we illustrate an auction-based resource allocation model.

\subsection{Different Game Models for D2D Direct Communication}

\begin{itemize}
 \item {\em Noncooperative power control game}: The transmit power of the D2D users must be properly controlled so as to coordinate the interference from D2D communication to a cellular network as well as the interference among D2D users themselves. Thus, interactions among the self-interested cellular and D2D users sharing the same channel and controlling the transmit power can be modeled as a noncooperative game. Specifically, the players are the D2D transmitters, the action is the transmit power of the D2D transmitter on each channel, and a utility (i.e., payoff) function can be defined as the negative of the total transmit power of the D2D transmitter. Multiple D2D users can reuse the same spectrum with cellular user. Thus, a power control method can be designed by iteratively updating the transmit power of D2D users in a self-organizing manner~\cite{Wang2013_ICC}.

 \item {\em Stackelberg-type game for local radio resource allocation}: With D2D communication underlaying cellular networks (i.e., D2D users reuse the licensed spectrum of the cellular networks), the concept of the Stackelberg game is well suited to analyze the radio resource allocation. In \cite{WangF2013}, a Stackelberg game is formulated and solved to obtain the transmission strategies of the cellular and D2D users. Given the hierarchical structure of the Stackelberg game, the players are cellular users (i.e., leaders) and D2D users (i.e., followers). A cellular user owns the channel and it can charge a D2D user some fees for accessing the channel. The fees are fictitious money to coordinate and control transmissions of the D2D user. Thus, the cellular user has an incentive to share the channel with the D2D user if it is profitable, and the cellular user has the right to decide the price. Given the charging price, the D2D user can choose the transmit power and channel to maximize its utility. The utility of the cellular user can be defined as its own throughput plus the revenue that it earns from the D2D user. The fee (i.e., price) should be decided according to the cellular user's (i.e., leader's) own consideration. For the D2D user as a follower, the utility is the difference between its throughput and the cost that it pays to the cellular user for using the channel. An equilibrium can be reached by both the leader and the follower.

\item {\em Auction-based model for radio resource allocation}: Auction theory is one branch of game theory which is widely used in trading if the price of a commodity and service is undetermined. There are many possible designs~(or sets of rules) for an auction. The typical issues in the auction include the efficiency of a given auction design, optimal and equilibrium bidding strategies, and revenue comparison. Some typical auctions include the Vickrey-Clarke-Groves~(VCG) auction, share auction, double auction, and combinatorial auction.


Specifically, the combinatorial auction~(CA)-based resource allocation mechanism allows an agent~(bidder) to place bids on combinations of resources, called ``{\em packages}'', rather than just individual resource block~(RB). The CAs motivate the bidders to fully express their preferences, which is beneficial for improving system efficiency and auction revenues. The auctioneer announces an initial price for each item, and then the bidders submit to the auctioneer their bids at the current price. As long as the demand exceeds the supply, or the supply exceeds the demand, the auctioneer updates~(raises or reduces) the corresponding price and the auction goes to the next round. The overall gain, which includes the total gain of the auctioneer and all bidders, does not depend on the pay price, but is equal to the sum of the allocated packages. Note however that pricing is not a typical aim in CA-based resource allocation mechanism, but the efficiently allocation. Pricing is used as a tool to achieve the maximum revenue for the auctioneer.

The CA game can be used for resource allocation among D2D links reusing the same cellular channels with the objective of optimizing the system capacity. In particular, with an iterative combinatorial auction~(ICA), the bidders submit multiple bids iteratively and the auctioneer computes the provisional allocations and asks the prices in each auction round~\cite{Xu2012}. In the following section, we will present a combinatorial auction-based approach for resource allocation for D2D direct communication.

\end{itemize}

Table~\ref{tab:smmary_game} summarizes the game theoretic approaches used for radio resource allocation of D2D direct communication.  Typically, noncooperative game approaches result in suboptimal solutions but have smaller signaling overhead in comparison to cooperative methods. Nevertheless, the auction game can obtain near optimal performance at the cost of certain signaling overhead between the D2D users and cellular networks.
\begin{table*}[h!]
\begin{center}
\caption{Summary of resource allocation game models for D2D direct communication}\label{tab:smmary_game}
\begin{tabular}{|m{30mm}|m{70mm}|m{50mm}|}
\hline \bf{Application} & \bf{Game model} & \bf{Solution}\\
\hline Power control & Noncooperative static game~\cite{Wang2013_ICC}: Cellular and D2D users as players & Power allocation for D2D or all users by iterative step update algorithm\\
\hline Local radio resource allocation & Stakelberg game~\cite{WangF2013}: Cellular user as leaders and D2D users as followers  & Channel allocation for D2D users by iterative best response algorithm\\
\hline Global radio resource allocation & Combinatorial auction~\cite{Xu2012}: Cellular networks as bidders and D2D users as auctioneer & Channel allocation for all users by iterative bid update algorithm\\
\hline
\end{tabular}
\end{center}
\end{table*}

\subsection{Combinatorial Auction-Based Resource Allocation}
The system consists of one eNB, multiple cellular users that receive signals from the eNB, and multiple D2D users using licensed spectrum resources. Since the D2D users share the same spectrum resources with cellular users, the cross-tier co-channel interference must be carefully controlled to optimize the network performance. Accordingly, appropriate channels allocated to the cellular users should be assigned to the D2D users to minimize the interference and achieve a higher system sum-rate. Basically, we treat the allocation of multiple D2D users with all cellular users' channels as the combinatorial auction-based resource allocation mechanism. The mechanism allows an agent (bidder), which is a cellular network, to place bids on combinations of resources rather than just individual RB to an auctioneer. \cite{Xu2012} proposes the reverse iterative combinatorial auction (R-I-CA) for channel allocation of D2D communication with multiple users. The key components of the combinatorial auction are
\begin{itemize}
 \item Auctioneers: Cellular and D2D users;
 \item Bidders: Cellular networks;
 \item Utility: The difference between the sum-rate that can achieve through the allocated channels and  the cost incurred due to signaling overhead.
\end{itemize}

In the auction, all the spectrum resources are considered as a set of RBs. The cellular networks as the bidders compete to obtain the demand (i.e., to sell their RBs) to the cellular and D2D users. In this case, the demand of the cellular network is considered as the ``package'' and will be auctioned off as the commodity in each auction round. The total spectrum is divided into RBs, each of which is allocated to one cellular user to transmit data. By the auction game, the spectrum units are also assigned to packages, with each package consisting of at least one mobile user. In other words, the cellular networks compete to obtain demand from mobile users for improving their own benefits. Although the cellular network as the bidder can gain revenue from mobile users, there exist some costs due to the transmission of  control signals  and overhead due to information feedback. Bidder's utility quantifies satisfaction of the bidder getting the package from mobiles, and the outcome of the auction is the spectrum allocation, which allocates a corresponding package to each bidder.

During the auction game, the mobile user as the auctioneer first announces an initial price for each RB. Then the bidders submit to the auctioneer their bids at the current price. As long as the demand exceeds the supply, or on the contrary the supply exceeds the demand, the auctioneer updates (raises or reduces) the corresponding price, and the auction proceeds to the next round. The allocation can be determined by multiple iterations, which continues until all the user links are auctioned off or every channel wins a package to reach the equilibrium solution. This combinatorial auction game has three important properties:
\begin{itemize}
 \item \emph{Convergence}: the number of iterations to reach the solution is finite.
 \item \emph{Price monotonicity}: the raising item prices in a round can reflect the competitive situation, which brings efficiency improvement.
 \item \emph{Complexity}: the complexity of the auction-based scheme is $\mathcal{O}(n\left( {2^m - 1} \right)+t)$, where the number of items to be allocated is $m$, the number of bidders is $n$, and the total number of iterations is $t$.
\end{itemize}

\begin{table}
\centering
\caption{Simulation parameters} \label{parameters}
\begin{tabular}{|p{5cm}|p{5cm}|}
 \hline Parameter & Value\\
 \hline
 \hline Cellular layout & Isolated cell, 1-sector\\
 \hline Cell radius & 500 m\\
 \hline Maximum D2D pair distance & 20 m \\
 \hline Cellular UE's transmit power $P^c$ & 23 dBm\\
 \hline D2D's transmit power $P^d$ & 23 dBm \\
 \hline Noise power & -104 dBm \\
 \hline Noise figure & 7 dB \\
 \hline Path-loss model & ITU UMi \\
 \hline Small-scale fading & Rayleigh fading coefficient with zero mean and unit variance \\
\hline
\end{tabular}
\end{table}

\begin{figure}[h!]
\centering
\includegraphics[width=4.5in]{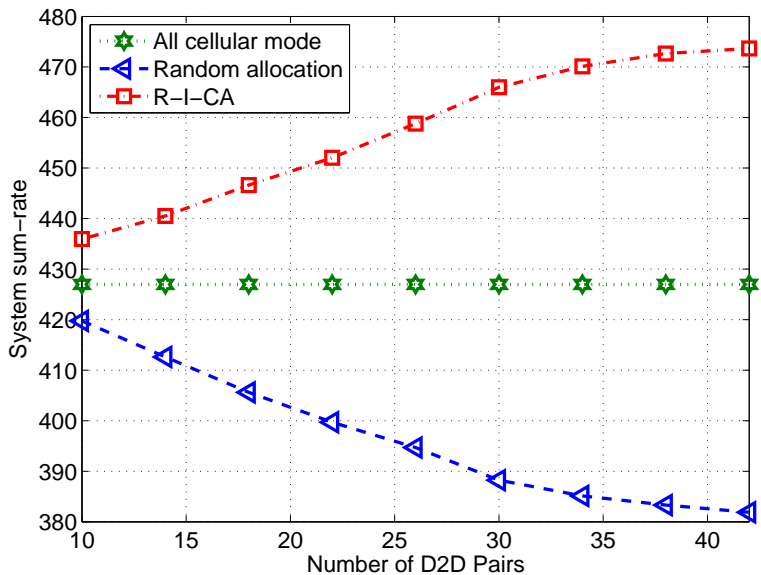}
\caption{System sum-rate for different allocation algorithms.} \label{D2D-D}
\end{figure}

For  a single cell environment Fig.~\ref{f:sysmodel} shows some simulation results based on the parameters in Table~\ref{parameters} to illustrate the performance of the proposed R-I-CA algorithm. Fig.~\ref{D2D-D} shows the system sum-rate with different numbers of D2D users using the combinatorial auction in comparison with random resource allocation and all cellular mode cases. Fig.~\ref{D2D-D} shows that when the number of D2D pair increases, the performance of the R-I-CA algorithm first increases before it is saturated. The reason is that a large number of D2D pairs will result in high co-tier interference as a result of which the bidders stop buying resources due to the unaffordable purchasing price.
Fig.~\ref{D2D-D} also shows that the random allocation is a worse option than making all the mobiles to adopt the traditional cellular mode (i.e., all the mobiles connect to the eNB), which may cause both co-tier and cross-tier interferences.

\section{Game Theory Models for D2D Local Area Networks}\label{sec:D2D LAN}

For resource allocation of group communication and multi-hop relay communication in D2D LANs, cooperative game models will be more suitable than the noncooperative approach. In the noncooperative approach, each mobile makes individual decisions, which may lead to severe interference. However, with a cooperative approach, the mobiles cooperate with each other to maximize its utility function, and consequently, a better network performance can be achieved.

In a cooperative game, players are able to make contracts which are mutually beneficial. The players cooperate to maximize a common objective of their coalition, and can coordinate strategies to agree on how the total payoff is to be divided among themselves. The coalitional game and Nash bargaining game are two major types of cooperative games. In the following, we will discuss how the coalitional games can be applied for resource allocation in D2D LANs. Also, we illustrate a coalitional game model for content distribution based on D2D communication.

\subsection{Coalition Formation Game Models for D2D Communication}

\begin{itemize}

\item {\em Coalition formation game for group communication}:

 In the coalition formation game, a set of players (i.e., D2D users) intend to form cooperative groups (i.e., coalitions). A coalition represents an agreement among the players to act as a single entity formed by players to gain a higher payoff, and the worth of this coalition is called a coalitional value. Two common forms of coalitional games are strategic form and partition form. In the strategic form, the value of a coalition depends on the members of that coalition only. In the partition form, the value of a coalition also depends on how the other players outside the coalition are structured. Coalitional game models can be developed with either transferable payoff or nontransferable payoff. In a transferable payoff coalitional game, utility serves like money and can be alloted to different players. In a nontransferable payoff coalitional game, different players have different interpretation of utilities, and the utilities cannot be distributed among players arbitrarily.

For the strategic-form coalitional formation games, a typical algorithm is the merge-and-split algorithm with the  two following operations:
\begin{itemize}
 \item \emph{Merge:} Coalitions merge to a single coalition whenever mutual benefits exist.
 \item \emph{Split:} A coalition splits whenever this splitting can provide better payoffs.
\end{itemize}
 It has been proved that the merge-and-split algorithm will always converge to a set of stable coalitions, in which no individual player has interests in changing coalition through a merge or split operation to achieve a higher utility.


\item \emph{Coalition graph game for multi-hop relay communication:} The concept of user relaying can be exploited to cope with the unreliable wireless links (due to both deep fading and co-channel interference) in a D2D LAN. A coalitional graph game can be used to model multi-hop D2D relaying. In this model, the utility functions of players are related to the specific graph that interconnects them to achieve specific objective, e.g., successful transmission rate or coverage extension. The mobiles as the relays form a directed tree graph to improve their utility by considering packet successful transmission rate by properly selecting the appropriate route and power level. In the game, the interactions can be represented by a directed graph $G(\mathcal{V},\mathcal{E})$ with $\mathcal{V}$ denoting the set of all nodes (the mobiles and eNBs) and $\mathcal{E}$ denoting the set of all edges (eNB-to-mobile links and D2D links). For any $i,j \in \mathcal{V}$, we say the link from $i$ to $j$ exists, if $e_{i,j} \in \mathcal{E}$. For any mobile $i \in \mathcal{N}$, only positive utility can be extracted from both the effective segments received from and transmitted to other mobiles, which depends on the current links associated with mobile $i$. Then, the utility function becomes a graph-based function, denoted by $U_i(G)$. Many algorithms, such as the myopic dynamics algorithm, can be applied for the coalitional graph game, which results in a directed graph that coordinates the transmissions in the network~\cite{Wang2013_tvt}.

 \item \emph{Overlapping coalition formation game and incentive mechanism for collaborative smartphone sensing:} The smartphones today not only provide simple communication services, but also can be used to sense and collect data from  the ambient environment. However, the quality and quantity of the sensed data uploaded by the users may not always satisfy the applications' requirements due to the lack of sufficient number of participating users. In order to encourage the mobiles to contribute their resources to the sensing tasks, cooperative games can be utilized in which D2D users can collaborate with each other to form coalitions and distribute their resources to different coalitions. Since each task can be accomplished by multiple users, and each user can be involved in multiple tasks,  the coalitions representing different sensing tasks may overlap with each other. Therefore, this is an overlapping coalition formation game~\cite{Di2013,Duan2012}. In the coalition-based smartphone sensing scheme, each user is seen as a player and the distributions of RBs in different coalitions are the possible strategies. Each coalition can be described by using both coalition members and RBs invested by each coalition member. The users are paid in certain forms by the coalitions based on their resource allocation.

\item {\em Matching and contract theory tools for mobile social D2D networks:} Performance of peer-to-peer communication in a D2D LAN can be improved by exploiting the social relationship of the D2D users. In this case, resource management methods need to jointly optimize a social network layer and D2D LAN layer to improve the network performance. Game theoretical tools from matching theory and contract theory, which can incorporate the characteristics of both layers for an optimization problem, can be used to develop optimal methods for resource allocation.

\end{itemize}

Table~\ref{tab:smmary_game_Lan} summarizes the different game approaches for D2D LAN.

\begin{table}[h!]
\begin{center}
\caption{Summary of Game-theoretic Methods in D2D LAN}\label{tab:smmary_game_Lan}
\begin{tabular}{|m{30mm}|m{70mm}|m{50mm}|}
\hline \bf{Application} & \bf{Game Model} & \bf{Solution}\\
\hline Group communication & Coalitional game~\cite{Wang2013}: Players are mobiles and eNB, and coalition value is the sum rate & Joint channel allocation and mobile selection by merge-and-split algorithm\\
\hline Multi-hop relay communication & Coalition graph game~\cite{Wang2013_tvt}: Vertexes are eNBs and mobiles, and edges include eNB-to-mobile links and D2D links & Power and route allocation by myopic dynamics algorithm \\
\hline Collaborative smartphone sensing & Overlapping coalition formation game~\cite{Di2013}: Coalition consists of coalition members (players) and the resource
units  & Physical resource allocation by convex optimization \\
\hline
Mobile social D2D network & Joint optimization for social network and D2D layers & Contract theory or matching theory \\
\hline
\end{tabular}
\end{center}
\end{table}

In the following section, we elaborate on a coalition graph-based resource allocation method for D2D LAN communication for a content distribution application.

\subsection{Coalitional Game-Based Resource Allocation for Efficient Content Distribution}

Driven by commercial interests, popular content distribution, as one of the key services in many hotspots such as stadium networks or concert networks, has recently received considerable attention. In this subsection, we focus on one most representative scenario: \emph{group communication}. Fig.~\ref{seed} presents a simple scenario to distribute a popular file to D2D LAN through traditional cellular networks. We introduce the use of a coalitional game for efficient content dissemination. In this scenario, $N$ users want the same file from the Internet, while only $K$ `\emph{seeds}' have already downloaded it. Instead of using more RBs to download the file directly from the eNB, the rest $N-K$ `\emph{normal}' UEs can ask the seeds to send the file using D2D communication. The performance of this approach is determined by which RBs are selected and with which D2D links they share their RBs~\cite{Wang2013}.

A coalition formation game can be employed in this scenario, which consists of the following components:
\begin{itemize}
 \item \emph{Players:} $N$ D2D UEs and $M$ cellular UEs.
 \item \emph{Coalition:} Each coalition contains one and only one cellular UE. The rest of members are the D2D UEs using the same RB.
 \item \emph{Coalition Value:} In a specific coalition, the seeds and the normal UEs form D2D links between each other. The coalition value is the sum rate of all the D2D links and the cellular link.
\end{itemize}
Since each coalition uses a different RB, the links in different coalitions do not interfere with each other, and thus, the coalition formation game has a strategic form. The algorithm for this game can be based on a `\emph{switch}' operation, which describes that a player may leave the current coalition and join a new coalition if the total value of both coalitions is strictly increased. The switch operation can be simply seen as a combination of a split and a merge. As the total value of the entire system is strictly increased by each switch operation, we can expect a set of stable coalitions in which no switch operation is preferred.

\begin{figure}[!t]
\centering
\includegraphics[width=4.5in]{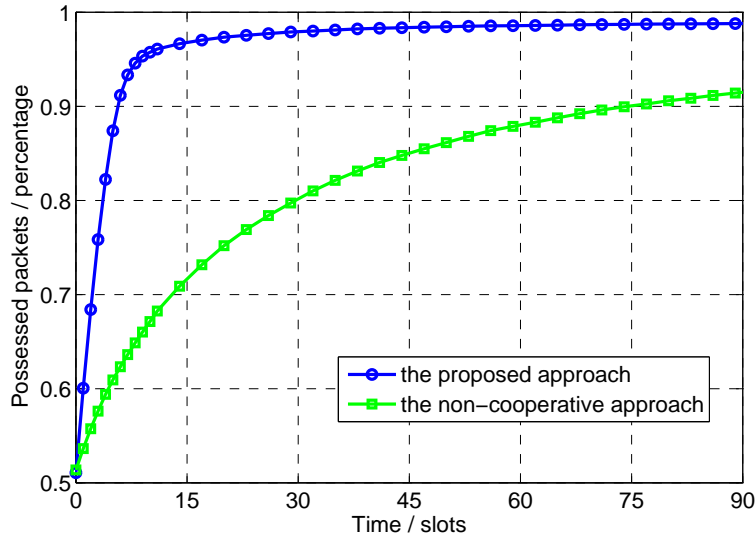}
\caption{Total possessed packets by the proposed coalition game-theoretic approach and the noncooperative approach.} \label{service_rate}
\end{figure}

In Fig.~\ref{service_rate}, it shows the cumulative service curves of both proposed coalition game-theoretic approach and noncooperative approach. First, we can see that the proposed approach performs much better than the noncooperative approach. In the noncooperative approach, each mobile makes individual decisions, which may lead to severe data collisions. However, in the proposed approach, the mobiles cooperate with each other to maximize the utility function, and consequently, the proposed approach achieves a better performance in service rate.

\section{Potential Research Directions in Designing Game Theoretic Methods for D2D Communication}

Game theoretical tools can be used to address many other resource allocation problems for D2D communication. Several potential research directions are outlined below.

\subsection{Device-to-Device Direct Communication}

\begin{itemize}

 \item {\em Cooperative game models for multi-cell optimization:} In practice, D2D communication may take place in multiple cells as the mobiles in each D2D pair can be associated with different eNBs. This can further save transmission time slots compared with single-cell case. In a single-cell case, D2D direct communication takes one time slot while traditional cellular communication needs two time slots. As a result, this requires proper coordination among several base stations or eNBs. Therefore, cooperative game models can be used to achieve multi-cell optimization and intercell interference coordination. Furthermore, additional time delay needs to be taken into account as an extra cost due to the signalling exchange among the eNBs.

 \item {\em Evolutionary game models for load balancing:} In D2D communication, each mobile can operate in two modes: cellular and D2D modes. In the cellular mode, the mobile communicates with an eNB. In the D2D mode, the mobile transfers data among each other directly. The users have choices to operate either in D2D or cellular mode. There are many factors related to  such mode selection including the distance between users, transmission quality to the eNB, interference conditions and performance requirements. Distributed selection of transmission modes of the D2D users (i.e., cellular mode or D2D communication mode) will be required for a scalable implementation of D2D networks. In this context, the use of evolutionary game models to balance the traffic for cellular and D2D users is worth studying.

 \item {\em Multi-level game models for cross-layer design:} For D2D communication, the cross-layer optimization problem of OFDMA networks becomes more complicated, where multiple D2D users are allowed to coexist with one cellular user as an underlay. Hence, how to effectively coordinate space, time, frequency, power, and device is a challenging issue. For joint allocation of multiple resources, multi-level game models can be developed. For instance, a noncooperative game model can be applied for power allocation, while in the upper level, an auction game can be used to assign radio resources.

 \item {\em Differential game models for energy efficiency:} One major benefit of short-range D2D communication is high energy efficiency, which can save battery power to increase handset life time, and can also detour the traffic from the eNBs to save their transmit power for green communication. To study the tradeoff between the total transmit power and system throughput, a differential game model can be readily utilized for analysis and design of a hybrid network.
\end{itemize}

\subsection{Device-to-Device Local Area Networks}

\begin{itemize}

\item {\em Incentive mechanism for cooperative D2D communication:} Cooperative D2D LAN communication can utilize a mobile as a helper relay to extend coverage, or improve spatial diversity through the information exchange between mobiles. Hence, it would be desirable to provide incentive for intermediate nodes to participate in the cooperation process, e.g., when mobiles are out of the service coverage but need other mobiles to forward information. In this case, a rewarding system regarding proper payment transfer should be developed to motive the collaboration among mobiles, and a contract game can be a good option.

\item {\em Matching game models for mobile crowdsourcing-based applications:} For crowdsourcing applications, collaborative D2D communication and network can obtain needed services, ideas, or content by soliciting contributions from an online community. Each contributor  adds a small portion of its own initiative to the greater result. Application examples such as multi-file sharing among mobiles, and collaborative indoor localization and navigation. Consequently, matching theory models can be adopted by assigning different mobiles to the corresponding interested targets, and coordinate the resource allocation process, such as to provide a mechanism to generate cost-effective alternatives to those expensive centralized solutions
\end{itemize}

\section{Conclusion}\label{sec:Summary}

Game theory has been used to model and analyze the noncooperative and cooperative behaviors of mobile users in the context of D2D communication underlaying cellular networks. The game theoretic models are useful for designing radio resource allocation algorithms to achieve stable and efficient solutions by allowing D2D users efficiently reuse licensed spectrum of cellular users. In this article, we have discussed different game models developed for allocating radio resources for \emph{D2D direct} and \emph{D2D LAN} scenarios. These models have been categorized based on the types of games. For D2D direct communication, noncooperative game and auction game models are suitable to solve the resource allocation problems. For D2D LANs, collaboration among mobiles is required, and thus, cooperative game models such as the coalition formation games can be used. We have presented an auction model and a coalitional game model in detail. We have also outlined potential research directions on developing game theoretic models to solve several important radio resource management problems for D2D communication.


\begin{thebibliography}{1}

\bibitem{Song2010}
L. Song and J. Shen, {\em Evolved Network Planning and Optimization for UMTS and LTE}. Auerbach Publications, CRC Press, 2010.

\bibitem{Doppler2009}
K. Doppler, M. Rinne, C. Wijting, C. Ribeiro, and K. Hugl, ``Device-to-device communication as an underlay to LTE-advanced networks," {\em IEEE Commun. Mag.}, vol. 47, no. 12, pp. 42--49, Dec. 2009.

\bibitem{3GPP}
3GPP TR $22.803$, v$12.01.0$, Mar. 2013.

\bibitem{WangF2013}
F. Wang, L. Song, Z. Han, Q. Zhao, and X. Wang, ``Joint scheduling and resource allocation for device-to-device underlay communication," {\em IEEE Wireless Communications and Networking Conference (WCNC)}, Shanghai, China, Apr. 2013.

\bibitem{Pei2013}
Y. Pei and Y. Liang, ``Resource allocation for device-to-device communications overlaying two-way cellular networks," {\em IEEE Transactions on Wireless Communications}, vol. 12, no. 7, pp. 3611--3621, Jul. 2013.

\bibitem{Yang2013}
M. J. Yang, S. Y. Lim, H. J. Park, and N. H. Park, ``Solving the data overload: Device-to-device bearer control architecture for cellular data offloading," {\em IEEE Vehicular Technology Magazine}, vol. 8, no. 1, pp. 31--39, Mar. 2013.

\bibitem{Phunchongharn2013}
P. Phunchongharn, E. Hossain, and D. I. Kim, ``Resource allocation for device-to-device communications underlaying LTE-advanced networks," {\em IEEE Wireless Communications}, vol. 20, no. 4, Aug. 2013.



\bibitem{Wang2013}
T. Wang, L. Song, and Z. Han, ``Popular content distribution in CR-VANETs with joint spectrum sensing and channel access," {\em IEEE Journal on Selected Areas in Communications}, vol. 30, no. 9, pp. 538--546, Sep. 2013.

\bibitem{Yu2011}
C.-H. Yu, K. Doppler, C. Ribeiro, and O. Tirkkonen, ``Resource sharing optimization for D2D communication underlaying cellular networks," {\em IEEE Trans. Wireless Commmun.}, vol. 10, no. 8, pp. 2752--2763, Aug. 2011.

\bibitem{Xu2012}
C. Xu, L. Song, Z. Han, D. Li, and B. Jiao, ``Resource allocation using a reverse iterative combinatorial auction for device-to-device underlay cellular networks," {\em IEEE Globe Communication Conference (Globecom)}, Los Angels, CA, Dec. 2012.

\bibitem{Janis_VTC}
P. Janis, V. Koivunen, C. Ribeiro, J. Korhonen, K. Doppler, and K. Hugl, ``Interference-aware resource allocation for device-to-device radio underlaying cellular networks," {\em IEEE Vehicular Technology Conference 2009-Spring}, Barcelona, Apr. 2009.

\bibitem{Yu_TWC}
C.-H. Yu, K. Doppler, C. Ribeiro, and O. Tirkkonen, ``Resource sharing optimization for D2D communication underlaying cellular networks," {\em IEEE Trans. Wireless Commmun.}, vol. 10, no. 8, pp. 2752--2763, Aug. 2011.

\bibitem{Han2011}
Z. Han, D. Niyato, W. Saad, T. Basar, and A. Hjoungnes, {\em Game Theory in Wireless and Communication Networks: Theory, Models and
Applications.} Cambridge, UK: Cambridge University Press, 2011.

\bibitem{Wang2013_ICC}
F. Wang, C. Xu, Q. Zhao. X. Wang, and Z. Han, ``Energy-aware resource allocation for device-to-device underlay communication," {\em IEEE International Conference on Communications (ICC)}, Budapest, Hungary, Jun., 2013.


\bibitem{Wang2013_tvt}
T. Wang, L. Song, and Z. Han, ``Coalitional graph games for popular content distribution in cognitive radio VANETs," {\em IEEE Transactions on Vehicular Technology, special issue on Graph Theory and Its Application in Vehicular Networking}, vol. 62, no. 8, pp. 4010--4019, Oct. 2013.

\bibitem{Di2013}
B. Di, T. Wang, L. Song, and Z. Han, ``Incentive mechanism for collaborative smartphone sensing using overlapping coalition formation games," {\em IEEE Globe Communication Conference (Globecom)}, Atlanta, USA, Dec. 2013.

\bibitem{Duan2012}
L. Duan, T. Kubo, K. Sugiyama, J. Huang, T. Hasegawa, and J. Walrand, ¡°Motivating Smartphone Collaboration in Data Acquisition and Distributed Computing,¡± appear in {\em IEEE Transactions on Mobile Computing}


\end{thebibliography}
\end{document}